\documentclass[acmsigecom]{acmtrans2m}

\usepackage{amsmath,amsfonts,amssymb,  mathrsfs,mathtools}
\usepackage{xspace}
\usepackage{dsfont}

\newdef{definition}[theorem]{Definition}
\newdef{remark}[theorem]{Remark}

\newcommand{\eg}{{\em e.g.,~\xspace}}

\def\R{\mathbb{R}}
\def\cI{{\cal I}}

\title[Algorithmic Delegated Choice]{Algorithmic Delegated Choice: \\ An Annotated Reading List}

\author[M. T. HAJIAGHAYI and S. SHIN]{MOHAMMAD T. HAJIAGHAYI \\ University of Maryland \and SUHO SHIN \\ University of Maryland}

\begin{abstract} 
    The problem of delegated choice has been of long interest in economics and recently on computer science.
    We overview a list of papers on delegated choice problem, from classic works to recent papers with algorithmic perspectives.
\end{abstract}
            
            

            
\begin{document}
            
\begin{bottomstuff} 
Authors' addresses: \texttt{hajiagha@umd.edu}, \texttt{suhoshin@umd.edu}
\end{bottomstuff}
            
\maketitle

\section{Introduction}
The \emph{delegated choice} problem is a fundamental model of principal-agent interaction with numerous real-world applications, capturing the tension when a decision maker (principal) delegates the role of decision making to an informed but self-interested agent.
The model has its roots in classic economic theory introduced by~\cite{holmstrom1978incentives}, and has since evolved into a rich interdisciplinary area spanning economics, computer science, and operations research.
It considers a scenario where the principal \emph{cannot} commit to a contingent monetary transfer, and thus the principal needs to commit to a mechanism that specifies the characteristics of the agent's proposals that she is willing to accept.
Such a model is particularly motivated by practical scenarios such as public regulators who can only accept or reject proposals from private sector, or investors relying on recommendations from financial advisors who are not contractually incentivized by the investor’s returns.

In its canonical form, the principal must select an action from a discrete (or often continuous) set $\Omega$, where each action $i \in \Omega$ has a pair of random utility values $(X_i, Y_i)$: one for the principal and one for the agent. 
Only the agent observes these values and proposes an action for selection, while the principal only knows the distribution from which these random values are drawn, introducing an information asymmetry. 
Once the agent observes the realizations, he sends a signal (\eg proposing an action) to the principal, who then makes the final decision.
The agent seeks to maximize his own utility rather than the principal’s, resulting in moral hazard. 
To mitigate this, the principal commits to a \emph{screening mechanism} that selectively accepts the proposed action based on its values.
For instance, an eligible set $E_i \subseteq \R^2$ can be announced to the agent so that the principal will only accept the proposed action $i$ if $(X_i,Y_i) \in E_i$.

Recent work has shifted from classical results on existence and structure of optimal mechanisms to more algorithmic and computational perspectives. This reading list collects key papers in this growing literature, highlighting recent developments in various aspects of the delegated choice problem, including its connections to the prophet inequality~\cite{krengel1977semiamarts}, Pandora’s box~\cite{weitzman1978optimal}, and broader stochastic optimization problems. 

Our goal is to introduce the core problem setup, present both classic and recent contributions, and illustrate how this line of work connects to various research areas in the EC community.
As such, this article is neither exhaustive nor comprehensive; we hope it will serve as a useful starting point for readers to grasp the central ideas and emerging directions in delegated choice problem.

\begin{enumerate}
    \item Bengt Rober Holmstr{\"o}m. On incentives and control in organizations, \emph{Ph.D. dissertation thesis}, 1978 \& Bengt Rober Holmstr{\"o}m, On the theory of delegation, \emph{Bayesian Models in Economic Theory}, 1984.
    \\\\
    Seminal work by~\cite{holmstrom1978incentives,holmstrom1984} provides the \emph{foundational framework} for delegation as an optimal screening mechanism. Instead of the discrete choice model described above, the principal delegates an optimization problem by choosing a screening set $A' \subseteq A$ from which the agent selects an action $a \in A'$ that yields payoffs $v(a,\theta)$ and $u(a,\theta)$ to the principal and agent, respectively, where $\theta$ is a realized state observed only by the agent. 
    Holmstrom characterizes conditions under which a single interval is optimal, and shows that the discretion given to the agent increases as his preferences align more closely with the principal's.
    
    
    \item  Mark Armstrong, John Vickers. A model of delegated project choice, \emph{Econometrica}, 2010.
    \\\\
    \cite{armstrong2010model} provides the first \emph{discrete choice} model, which serves as a foundation for subsequent works. There are $n$ available actions,\footnote{In fact, they consider a setting where $n$ is also drawn from a known distribution.} each with a type $(u,v)$ drawn i.i.d. from a distribution, where the agent receives payoff $v$ and the principal receives $v + \alpha u$ for some $\alpha \ge 0$. Only the proposed action's type is verifiable, so the principal's goal is to design a screening mechanism over admissible types $(u,v)$, and the authors identify optimal mechanisms under specific payoff assumptions.
    
    
    \item Jon Kleinberg, Robert Kleinberg. Delegated search approximates efficient search, \emph{Proceedings of the 2018 ACM Conference on Economics and Computation (EC)}, 2018.
    \\\\
    \cite{kleinberg2018delegated} studies the approximate efficiency of the optimal delegation mechanism with respect to the \emph{first-best} benchmark, where the principal observes the utilities of all actions in hindsight and can choose what she wants, building on the model of~\cite{armstrong2010model}.\footnote{They also consider a search problem where the agent undergoes a costly search process to seek solutions, with connection to Pandora's box problem~\cite{weitzman1978optimal}.} It considers precisely the problem setup described in the introduction, and uncovers a surprising connection to the classical \emph{prophet inequality}~\cite{krengel1977semiamarts}. Specifically, they prove an \emph{equivalence} between the delegated choice problem and a version of prophet inequality with oblivious stopping rules. This leads to several delegation gap bounds, including a $1/2$-approximation with a threshold mechanism in the general case and a $(1 - 1/e)$-approximation in the i.i.d. case.
    
    
    \item Kiarash Banihashem, Mohammad T. Hajiaghayi, Piotr Krysta, Suho Shin. Delegated choice with combinatorial constraints, \emph{Proceedings of the 2025 ACM Conference on Economics and Computation (EC)}, 2025.
    \\\\
    \cite{banihashem2025} considers a natural follow-up to~\cite{kleinberg2018delegated}, asking to what extent the connection between delegated choice and prophet inequality carries over. They study a multi-choice setting with a universe $U$ of $n$ actions and a family of feasible sets $\cI \subset 2^U$, where the principal aims to select a set $I \in \cI$ to maximize $X_I = \sum_{i \in I} X_i$. Notably, they provide the first \emph{provable separation} between the two problems by showing that delegated choice under downward-closed constraints allows constant-factor approximation, whereas prophet inequality does not~\cite{rubinstein2016beyond}; they further show that the correspondence between the two problems holds \emph{if and only if} the constraint is a matroid.
    

    \item Curtis Bechtel, Shaddin Dughmi. Delegated stochastic probing, \emph{12th Innovations in Theoretical Computer Science (ITCS)}, 2021.
    \\\\
    \cite{bechtel2021delegated} proposes a delegated stochastic probing problem where the agent probes actions under an outer constraint $\cI_{out}$ and proposes a feasible set under an inner constraint $\cI_{in}$. Without an outer constraint or probing cost, their model coincides with~\cite{banihashem2025}, and they show that a greedy prophet inequality strategy against an almighty adversary, who observes every random bit of the algorithms and environments and decides a worst-case instance, can be implemented in this setting. 
    This yields several immediate corollaries on matroid, matching, and knapsack constraints via greedy online contention resolution schemes~\cite{feldman2016online}.\footnote{We note here that this does not contradict the necessity of matroid results shown in~\cite{banihashem2025}, since \cite{bechtel2021delegated} guarantees only that the principal’s utility exceeds the prophet’s, not an exact equivalence between the agent's choice and the prophet inequality algorithm's decision as shown by~\cite{banihashem2025}.}


    \item  Ali Kohdabakhsh, Emmanouil Pountourakis, Samuel Taggart. Simple delegated choice, \emph{Proceedings of the 2024  Annual ACM-SIAM Symposium on Discrete Algorithms (SODA)}, 2024.
    \\\\ 
    The assumption that the proposed action's utilities are easily verifiable is admissible as misreporting could be verifiable by implementing the action or it may incur a reputation effect.
    On the other hand, this could often be impractical when the delegation happens as a one-off interaction or if it entails an expensive cost of verification such as the delegated decision of a governmental policy.
    \cite{khodabakhsh2024simple} considers a mechanism that \emph{does not depend} on the utilities of the proposed values, but the principal completely rules actions in or out based on the distributional knowledge.
    They show that competing with the first-best benchmark is hopeless, and that the problem of computing the optimal mechanism is NP-hard, which is complemented by their $1/3$ approximate deterministic mechanism.
    
    \item Mohammad T. Hajiaghayi, Piotr Krysta, Mohammad Mahdavi, Suho Shin. Delegation with costly inspection, \emph{Proceedings of the 2025 ACM Conference on Economics and Computation (EC)}, 2025.  
    \\\\ 
    \cite{hajiaghayi2025} directly addresses the verifiability assumption by allowing the principal to \emph{inspect} the proposed action, and possibly others, at deterministic costs $c_i$, to verify utilities. In their extension of~\cite{kleinberg2018delegated}, the agent may misreport if inspection is unlikely, and delegation itself incurs a fixed cost $c_{del}$. 
    This model generalizes the Pandora's box problem with nonobligatory inspection~\cite{doval2018whether}, inheriting its NP-hardness~\cite{fu2023pandora,beyhaghi2023pandora}, and they show that while the first-best benchmark cannot be approximated, constant-factor approximate mechanisms exist in both costless and costly delegation settings when the cost of delegation is high or low.
    
    
    \item Suho Shin, Keivan Rezaei, Mohammad T. Hajiaghayi. Delegating to multiple agents, \emph{Proceedings of the 2023 ACM Conference on Economics and Computation (EC)}, 2023.  
    \\\\
    While the preceding works focus on Bayesian mechanisms, a few have explored prior-independent mechanisms in relaxed settings.\footnote{In the standard setup with a single agent, one can easily see that no prior-independent mechanism can approximate the first-best benchmark.} \cite{shin2023delegating} studies a multi-agent delegated choice problem where each agent proposes an action, but only the selected agent receives nonzero utility, introducing competition that benefits the principal. They consider both Bayesian and prior-independent mechanisms and show that a constant-factor prior-independent mechanism exists in the complete information setting with symmetric agents, whereas the benefit of having multiple agents depends heavily on the agents’ information and symmetry.
    
    \item Curtis Bechtel, Shaddin Dughmi. Efficient multi-agent delegated search, \emph{Proceedings of the 24th  International Conference on Autonomous Agents and Multiagent Systems (AAMAS)}, 2025.
    \\\\
    \cite{bechtel2025efficient} improves the results for Bayesian mechanisms under the incomplete information setting introduced by~\cite{shin2023delegating}. They achieve an approximation factor tending to $1$ as the number of agents increases, when the agents have symmetric sets of actions that are not necessarily i.i.d.
    This strengthens the approximation factor and relaxes the constraints introduced by~\cite{shin2023delegating}.
    
    \item Mohammad T. Hajiaghayi, Mohammad Mahdavi, Keivan Rezaei, Suho Shin. Regret analysis of repeated delegated choice, \emph{Proceedings of the Thirty-Eighth AAAI Conference on Artificial Intelligence (AAAI)}, 2024.
    \\\\
    Distributional knowledge, in practice, is usually constructed from historical data. \cite{hajiaghayi2024regret} considers a variant with prior-independent mechanisms where the principal can repeatedly interact with the agent to construct estimates of the distributions.
    They frame their setup as a stochastic multi-armed bandit problem, and propose no-regret learning algorithms for myopic and farsighted agents under the Lipschitzness assumption of the utilities, using tools from bandits with perturbed outputs.
\end{enumerate}

\bibliographystyle{acmtrans}
\bibliography{ref}

\begin{received}
\end{received}
\end{document}